# Imaging with two spiral diffracting elements intermediated by a pinhole

**José J. Lunazzi, Noemí I. R. Rivera and Daniel S. F. Magalhães**

*Universidade Estadual de Campinas, Instituto de Física Gleb Wataghin, Caixa Postal 6165, 13083-970 Campinas, SP, Brazil*

Pseudoscopic (inverted depth) images made with spiral diffracting elements intermediated by a pinhole is explained by its symmetry properties. The whole process is made under common white light illumination and allows the projection of images. The analysis of this projection demonstrates that the images of two objects pointing away longitudinally have the main features of standard pseudoscopic image points. An orthoscopic (normal depth) image has also been obtained with the breaking of the symmetry conditions.

# 1. Introduction

Images obtained using three diffractive elements were studied and discussed theoretically by Sweatt [1] where it was proposed that the configuration could be used for a lightweight telescope when coupled with a converging lens. Weingärtner [2] used a system composed of a diffraction grating and a holographic lens to obtain an image of a far object. A photograph showed a sharp image at 10 nm bandwidth and a rough image when using the visible spectrum. To our knowledge no experimental result was reported with a three diffractive elements image system.

Lunazzi [3-5] previously reported that the pseudoscopic image obtained with diffraction gratings intermediated by a slit converges only in the horizontal plane thus generating an image which is not astigmatic for visual observation but astigmatism precludes to obtain a convergent image. Reducing the slit to a pinhole eliminates astigmatism but reduces luminosity. Further work has shown [6, 7] that convergence of all rays at the image position is obtained with the same geometry with bidimensionally structured diffracting elements and with the replacement of the slit by a pinhole. The chosen DOE was a spiral one that come be considered equivalent to a circular one of constant period [8]. The element constitutes a diffractive axicon [9, 10] making as succession of convergent beams along a resulting luminous line. This convergence restores luminosity to the image projection because concentrates more light at the pinhole position. It was also reported [11] that an orthoscopic image can be obtained.

In this work, the above geometries are analyzed by ray tracing, and focusing experiments are performed. In Section 2 we present the basic formalisms required to describe pseudoscopic and orthoscopic images obtained with two diffracting elements intermediated by a pinhole. The experimental set up for focusing experiments is described in Section 3; the results and discussion are in Section 4.

## 2. Description

### *2.1 Pseudoscopic Image*

The diffracting elements we used to observe pseudoscopic images have the spiral groove format because digital data discs perfectly fulfill the requirements. The generation of non-diffracting beams by these elements was previously studied by Ferrari et al. [9].

Figure 1 shows the general view of the optical system used to obtain the pseudoscopic image of an object. To explain the imaging properties it suffices to consider that for each ray exiting from a point object a plane can be selected to determine the image position, which is perfectly symmetric. With this geometry, the existence of light rays can be assured within an extensive region reached by the diffracted light after the first element. As it can be seen in Figure 1, circular symmetry extends the above properties to a family of planes within the whole volume defined by the selected experimental geometry.

We can choose one of this planes for the analysis. Figure 2 shows two object points A and B placed in front of the first diffraction element DOE1. The diffracted light from DOE1 passes through the pinhole P and reaches the identical diffraction element DOE2. In the present analysis the diffraction effects due to the pinhole itself are neglected. The dotted line in Fig. 2 gives the alignment of the DOE1 and DOE2 spiral centers with the pinhole P. It can be seen in this figure that with respect to the pinhole plane the images A' and B' are symmetrical to A and B.

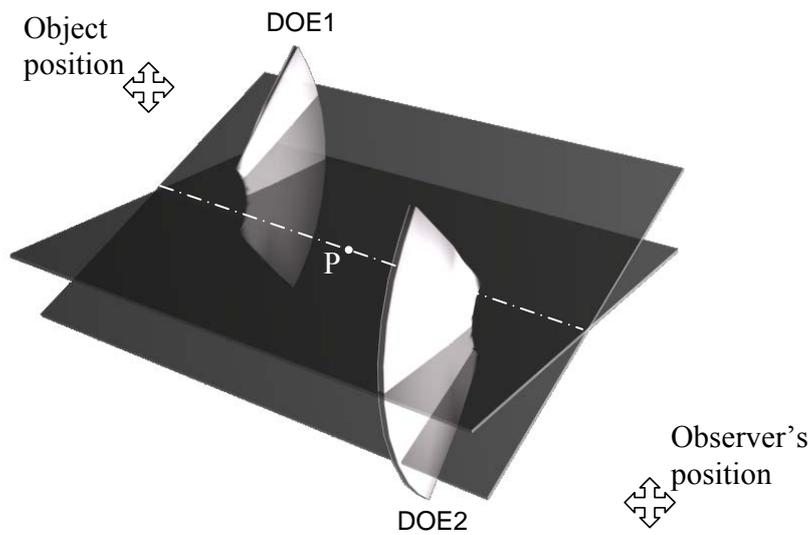

Figure 1. A general view of the optical system and the decomposing of the imaging process through a family of planes corresponding to circular symmetry.

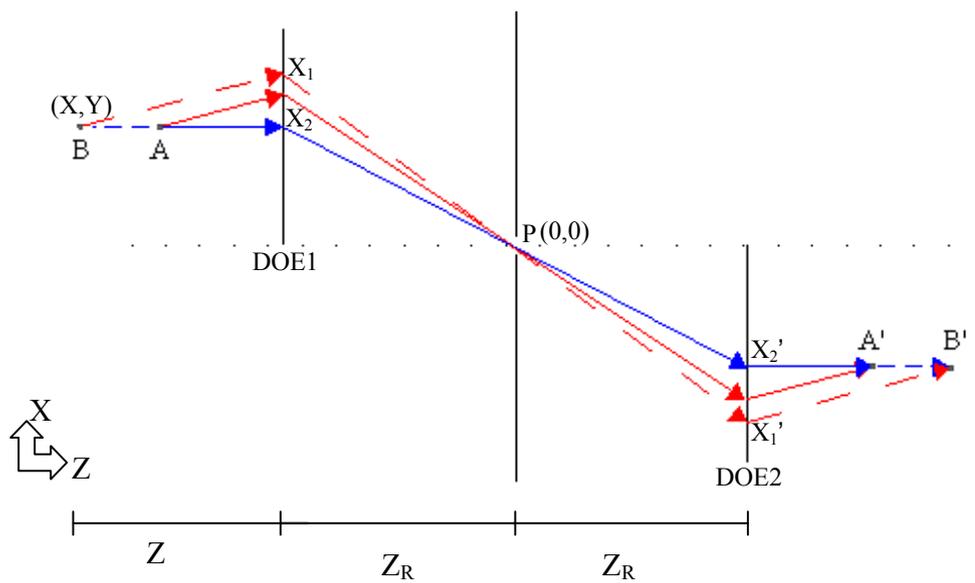

Figure 2. Image formation of two object points A and B.

The whole space can thus be decomposed in a set of oblique planes containing the symmetry axis. The spreading of the diffracted beam that the first diffractive element generates warrants the existence a ray through the pinhole P. The subsequent diffracted beam due to the second diffraction grating is directed towards the image point. The observer's position can receive rays from only one plane of the kind we selected, that of the plane where it is included. Rays not included in the selected planes travel through conical diffraction [12, 13]. Because rays spreading in different directions are present is that the observation of the whole image is possible from a single point of view. The observer sees the image of points at different planes because of the rays which are not on the planes we selected. With this geometry, the observer sees a convergent image resulting from first and second diffractions on the same side of its own position, satisfying the general grating equation,

$$\sin\theta_i - \sin\theta_d = \frac{\lambda \nu}{\sin\phi}, \qquad (1)$$

where $\theta_i$ represents the angle of incidence of light traveling from point A to the first diffracting element, $\theta_d$ represents the angle of diffraction for light that travels from points on the diffracting element to point P, $\phi$ is the angle a ray makes with the tangent to the grooves at the incidence point in case it goes out of the plane of Fig. 2. $\nu$ represents the inverse of the grating period, and an specific wavelength value, $\lambda$, is assigned to each ray. It suffices to characterize the image properties to consider ray tracing for rays with $\phi = 90°$. All other rays converge to the same image point because of symmetry.

As it can be expected for white light phenomena, extreme wavelength values are chosen to characterize the ray tracing for all other rays fall within these rays. For each plane, the light path for the first diffraction can be represented by the set of equations (2) and (3) that describes the trajectories as if the gratings were straight-line gratings:

$$\frac{X_1 - X}{\sqrt{(X_1 - X)^2 + Z^2}} + \frac{X_1}{\sqrt{X_1^2 + Z_R^2}} = \lambda_v \nu \; ; \quad \frac{X_2 - X}{\sqrt{(X_2 - X)^2 + Z^2}} + \frac{X_2}{\sqrt{X_2^2 + Z_R^2}} = \lambda_a \nu \quad (2)$$

Path calculations obtained after the second diffraction are on the set of Eqs. (3), from which one obtains the position of an image point $(X_i, Z_i)$ of an object point $(X, Z)$:

$$\frac{X_1'}{\sqrt{X_1'^2 + Z_R^2}} + \frac{X_i - X_1'}{\sqrt{(X_i - X_1')^2 + Z_i^2}} = -\lambda_v \nu \; ; \quad \frac{X_2'}{\sqrt{X_2'^2 + Z_R^2}} + \frac{X_i - X_2'}{\sqrt{(X_i - X_2')^2 + Z_i^2}} = -\lambda_a \nu \quad (3)$$

In Eqs. (2) and (3), the ordinates $X_1$ and $X_2$ are points at the first grating where the paths for blue and red light wavelengths are considered, respectively. Z and $Z_R$ are the lengths between the object and EOD1 and between EOD1 and the pinhole, respectively (as shown in Fig.2). The difference in the signal of the $\sin \theta_d$ in equation (1) and the coordinates of equations (2) and (3) follows the convention adopted by [14].

## 2.2 Orthoscopic Image

With the above described geometry, an orthoscopic image can be seen when flipping the second DOE and the changing the angle for observation. The angle is changed to select the opposite diffraction order. It should be noted that the system geometry remains identical to the one described above to observe a pseudoscopic image. The ray scheme for this situation is shown in Figure 3.

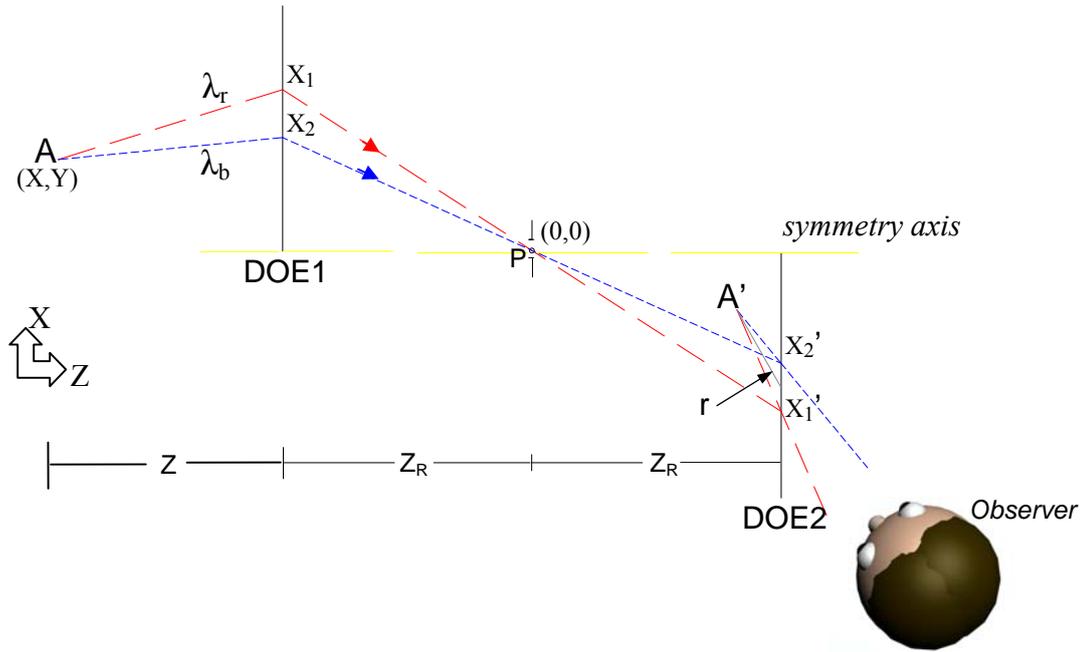

Figure 3. Ray-tracing for the image of a point white-light object.

In Fig. 3, the white-light object A shines the first of the two diffracting elements that are symmetrically located with respect to the pinhole P. The observer looks to a diverging image resulting from these diffractions located in between the pinhole and DOE2.

As it was the case in section 2.1 the equation (2) is valid. The paths at the second diffraction satisfy the equation (4), only differing by a signal from equation (3), from which one obtains the position of the image point:

$$\frac{X_1^{'}}{\sqrt{X_1^{'2}+Z_R^2}}+\frac{X_i-X_1^{'}}{\sqrt{(X_i-X_1^{'})^2+Z_i^2}}=\lambda_v v \; ; \qquad \frac{X_2^{'}}{\sqrt{X_2^{'2}+Z_R^2}}+\frac{X_i-X_2^{'}}{\sqrt{(X_i-X_2^{'})^2+Z_i^2}}=\lambda_a v \quad (4)$$

Figure 4 explains normal depth on the image for two object points A and B, located at different depth positions. As the figure shows, it was also verified that a shortening of the dimensions on the image happens in one direction along this plane. A left-to-right inversion is also a characteristic consequence of this image formation. Numerical calculations indicates that the image can be viewed binocularly, resembling a holographic image but with a spectral sequence of color change due to the wavelength changes associated with the displacements of the observer point of view. The image features obtained in the present study with circular gratings are already present in the case of the orthoscopic images obtained with ordinary straight-line gratings [5]. It is interesting to note that for both kinds of images, the diffracting elements can be anyone within a more general variety: the major requirements are that the two diffractive elements are identical ones and that the second element is symmetrically oriented with respect to the line normal to the pinhole, satisfying the above described ray-tracing conditions. The more regular the element, the more easy the alignment would be. A diffractive or holographic lens pair could bring more luminosity by concentrating the light at the pinhole region.

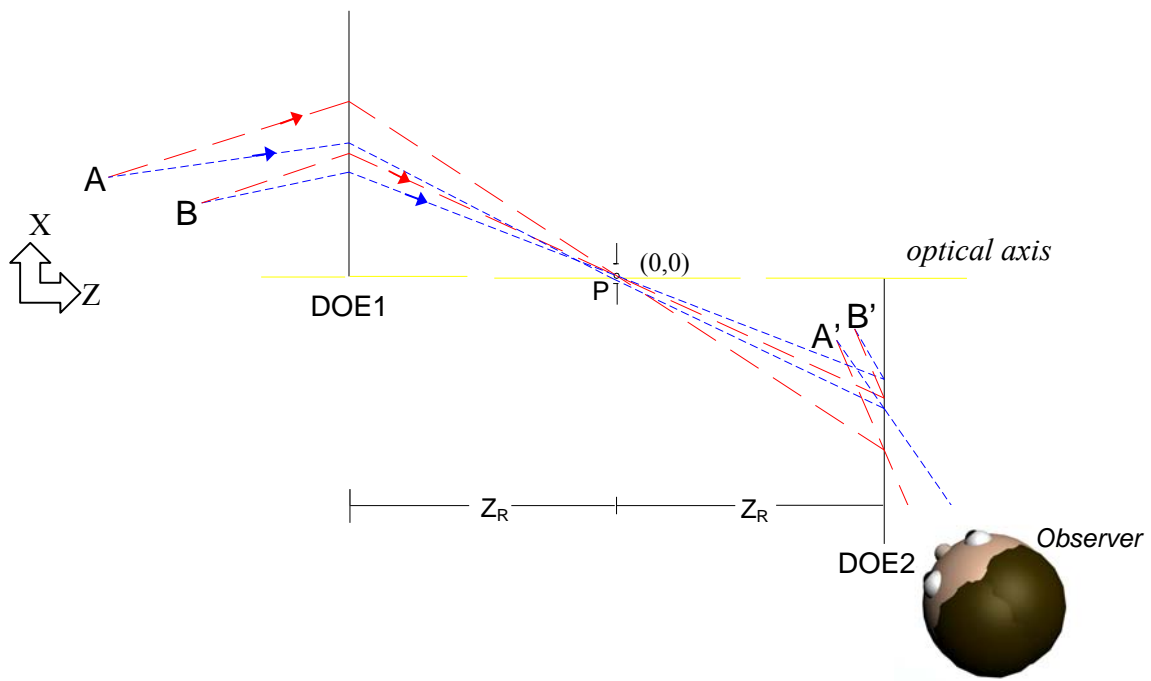

Figure 4. Ray-tracing scheme for the normal depth image.

Rays outside our set of selected planes can only reach the pinhole through conical diffraction. The symmetry condition does not applies anymore and this may be the cause of some aberration, an interest case for further study.

In Figure 5, there are four objects of height h in pairs at distances $Z_1$ and $Z_2$. The X coordinates of the objects named A or B is $X_A$ or $X_B$, respectively. Two planes (PI and PII) are shown characterized because they are common to the symmetry axis and includes the extremes of the objects $B_1$ and $B_2$. α is the angle between these planes, satisfying the following relationship:

$$\alpha = \text{ArcTan}\left(\frac{h}{x}\right) \tag{5}$$

These planes determine the extension of the object B1 and also of its image B1'. Shown at the figure are the paths of a single ray exiting from the extremes of B1 until they reach its image.

The light path can be obtained from the Eqs. (6) and (7):

$$\frac{\frac{X_{1,2}}{\cos\alpha} - \sqrt{X^2 + h^2}}{\sqrt{(\frac{X_{1,2}}{\cos\alpha} - \sqrt{X^2 + h^2})^2 + Z^2}} + \frac{\frac{X_{1,2}}{\cos\alpha}}{\sqrt{(\frac{X_{1,2}}{\cos\alpha})^2 + Z_R^2}} = \lambda_{a,v} \, v \tag{6}$$

$$\frac{-\frac{X_{1,2}}{\cos\alpha}}{\sqrt{(\frac{X_{1,2}}{\cos\alpha})^2 + Z_R^2}} + \frac{\frac{X_{1,2}}{\cos\alpha} - X_i}{\sqrt{(\frac{X_{1,2}}{\cos\alpha} - X_i)^2 + Z_i^2}} = \lambda_{a,v} \, v \tag{7}$$

The extension of the image B' is given by:

$$h' = X_i.\text{sen }\alpha \tag{8}$$

where Xi is determined by the system of equations (6) and (7).These equations are necessary to completely describe the imaging process.

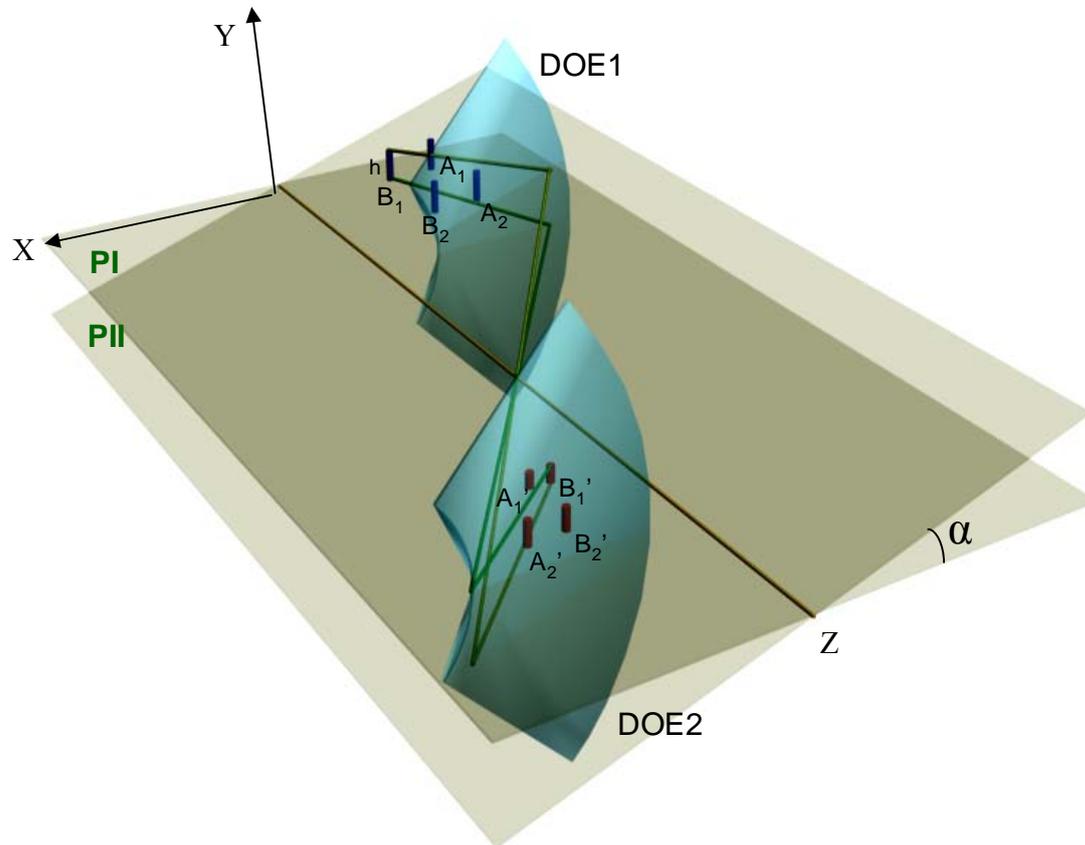

Figure 5. $B_1$' e $B_2$' are images of the objects $B_1$ e $B_2$ , all of them belonging to planes PI and PII.

## 3. Experimental setup

The system is shown in Figure 6. The spiral transmission gratings DOE1 and DOE2 are two pieces of a compact disk with spatial frequency of $658 \pm 5$ lines/mm at any radial distance, generating phase diffraction in the first order with 10.5 % efficiency. The elements used in this study were originally transmissive in order to reduce the influence of surface distortion, but reflective elements can often be made transmissive by removing its reflective coating.

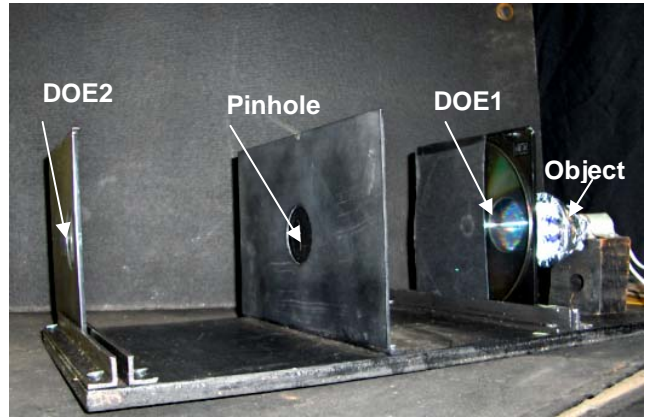

Figure 6. Experimental setup of a double diffraction system with a pinhole.

The objects employed have high brilliance to facilitate their locations and for easily registering images. These objects were filament lamps with transparent bulbs – halogeneous or tungsten. A video camera SONY HANDYCAM CCD-TRV57 with adjustable focus, linked through a webcam Intel CS430 acting exclusively as a capture device, was employed for registering the images on a computer.

Figure 7 is a photograph showing the light distribution generated by the white point object O (a white light LED) diffracted at DOE1 and reaching a region around the pinhole P. This region of diffracted light is 14 cm wide but only 4.5 cm in depth, corresponding to the first diffraction order selected in the present study. The dashed line identified as O.P. in Fig. 7, corresponds to the plane previously described in Fig. 2 on which the rays can pass the pinhole P, for the subsequent formation of the image.

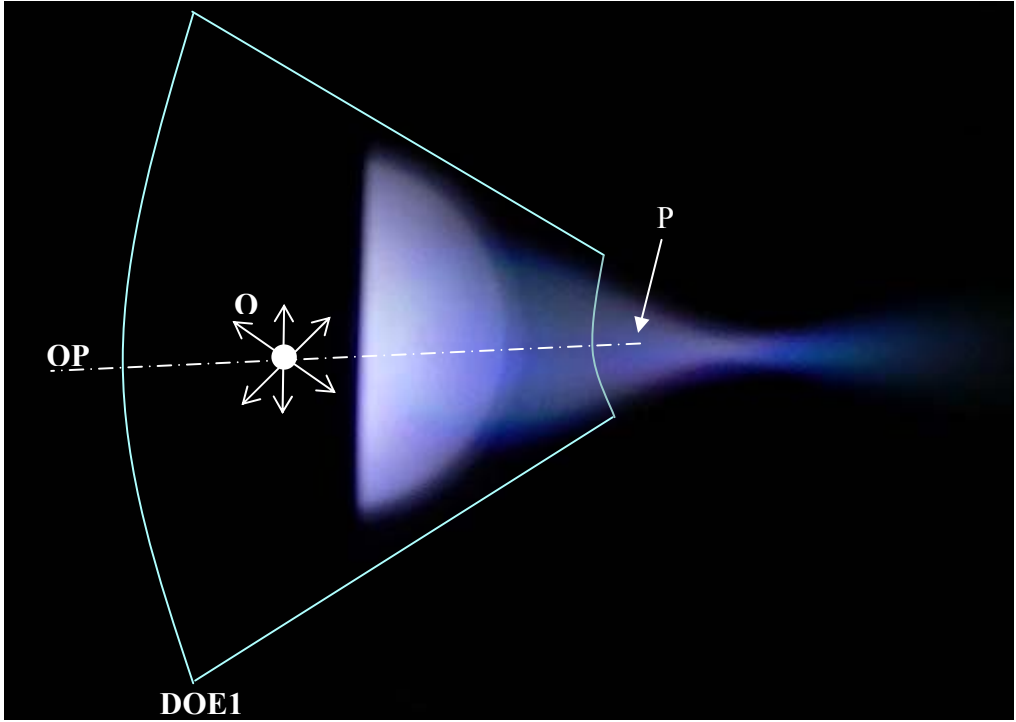

Figure 7. Light distribution reaching the pinhole P.

## 4. Results and Discussions

### *4.1 Pseudoscopic Image*

Figure 8 shows two tungsten filaments ($F_1$ and $F_2$) of 40 W lamps used as objects in this work. The filament directions are roughly orthogonal, $F_1$ positioned ahead of $F_2$, to the observer (camera) situated as shown in Figure 9 – this figure also gives an overall scheme of the experimental set up. The values of the main ordinates and distances identified in Fig. 9 are given in Table I. The image of the joint object constituted by the elements $F_2$' e $F_1$' is therefore pseudoscopic. The focalized images of each filament are shown in Figures 10 and 11. It should be noted that these images are inverted along the up-down direction. The spectral dispersion crossing the filament image $F_2^{'}$ (Fig. 10, right) are due to rays that had composed $F_1'$. The defocused points appear with a chromatic spectrum whose width is proportional to the distance to the symmetry plane. The width of this dispersion keeps the information of the depth of the image – known as "diffraction depth codification" [15,16].

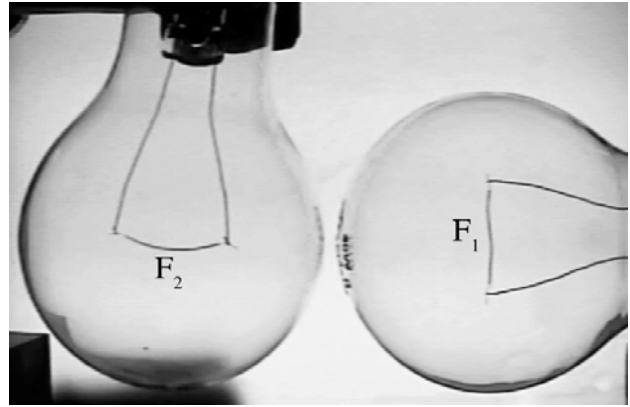

Figure 8. Picture of the objects $F_1$ and $F_2$ used to obtain the pseudoscopic image.

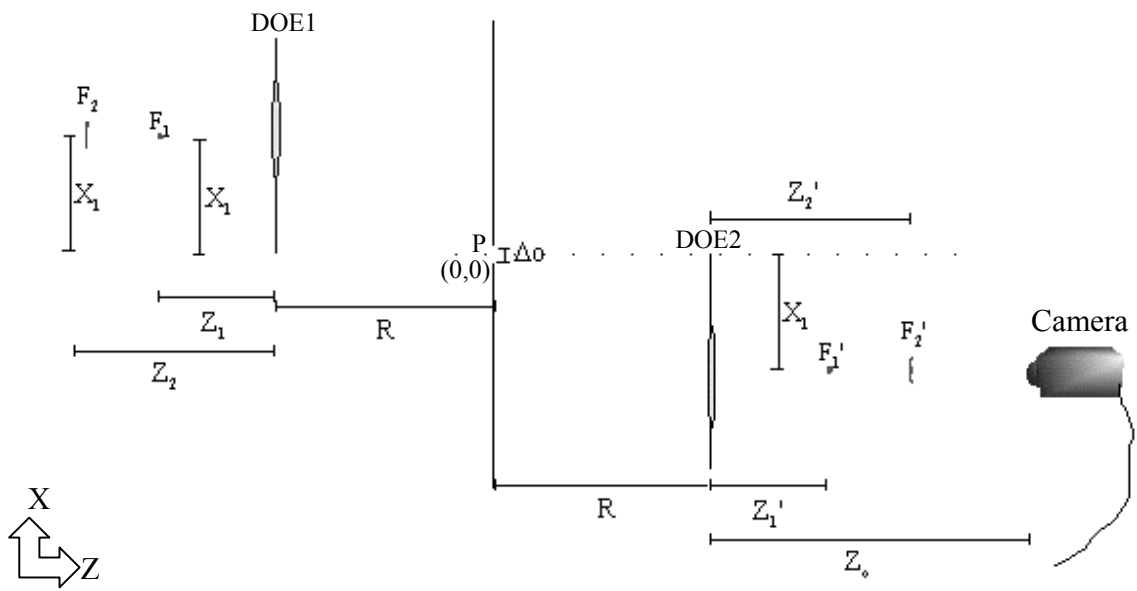

Figure 9. Aerial vision of the apparatus of image formation.

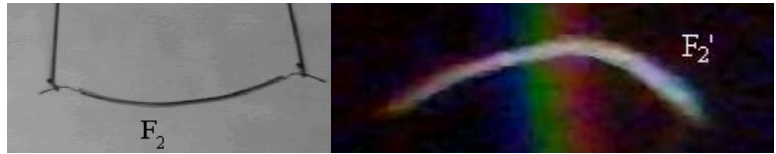

Figure 10. $F_2$ at left. The image of $F_2$ ($F_2'$) at right. The defocus of the image of $F_1$ are viewed.

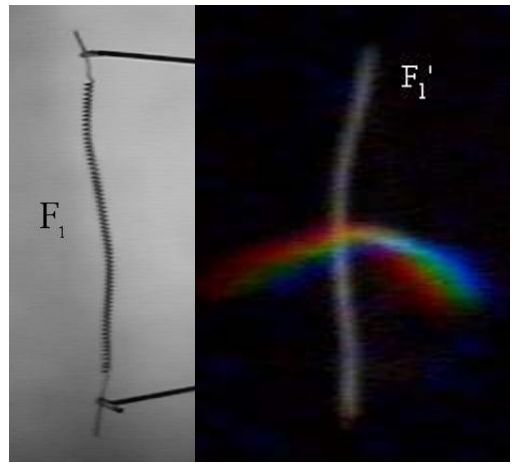

Figure 11. $F_1$ at left. The image of $F_1$ ($F_1'$) at right. The defocus of the image of $F_2$ are viewed as a spectral dispersion.

Using binocular observations to study image depth properties is not possible in the pseudoscopic case due to the restricted angular extension of the image. If the present set up had binocular capability, an observer placed in the position $Z_0$ of the camera would have seen $F_2'$ in front of $F_1'$. This is not the case in the present experiment.

In Table I, one can see that $Z_2' \cong Z_2$ and $Z_1' \cong Z_1$, indicating that the symmetry proposal for the model geometry is experimentally fulfilled.

Table I: Experimental checking of the pseudoscopic image

| R (mm) | $Z_o$ (mm) | $\Delta o$ (mm) | $X_1$ (mm) | $Z_1$ (mm) | $Z_2$ (mm) | $Z_1'$ (mm) | $Z_2'$ (mm) |
|---|---|---|---|---|---|---|---|
| 130 ± 1 | 477 ± 1 | 0.5 ± 0.1 | 35 ± 1 | 73 ± 2 | 122 ± 2 | 67 ± 3 | 117 ± 3 |

Table I: Experimental checking of the pseudoscopic image.

## *4.2 Orthoscopic Image*

In the present experimental set up, the observation of orthoscopic images was done with a white-light object: a 20 W halogeneous lamp with parabolic reflector. This device provided observations with a very bright object and a fraction of its diffracted light from the first element illuminates the region around the pinhole. Figure 12 (a) shows a photography of the object. Fig. 12 (b) shows the orthoscopic image obtained at an observer's distance $Z_c = (150 ± 5)$ mm. This picture shows that the horizontal dimension appears reduced as compared to the vertical, what is in agreement to our ray tracing analysis (section 3.2.3 below). At object distances close to the first diffracting element this reduction is not noticeable – it only appears when the object is moved away.

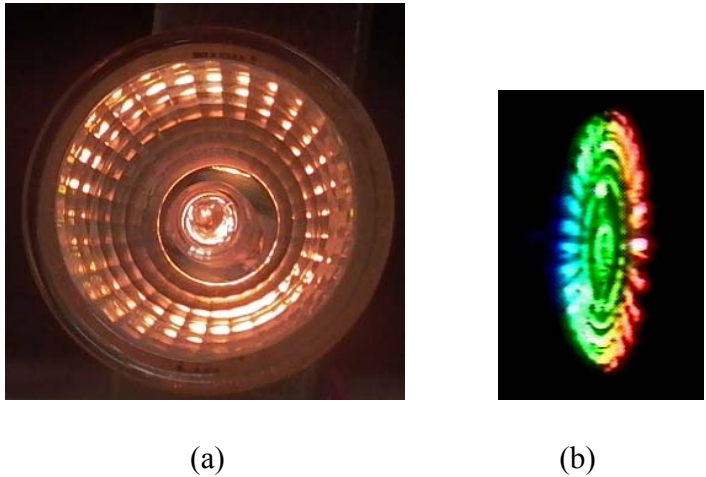

(a)            (b)

Figure 12. a) Photography of the halogeneous lamp employed as object. b) Orthoscopic image of the object obtained in our system.

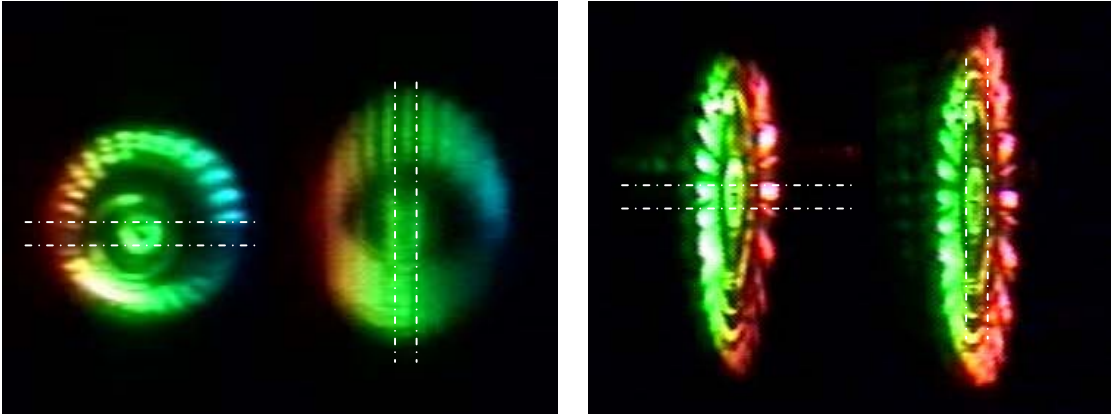

(a) (b)

Figure 13. The dashed lines represent the slit position in: a) double-diffracted images for the plane gratings case. b) double-diffracted images for the circular gratings case.

*4.2.1 Astigmatism of the image*

The observed astigmatism is attenuated, if not eliminated, as compared to the case of a diffraction grating pair [5] because the chosen diffractive elements act upon two dimensions thus forming a richer image. Figure 13 (a) shows two images of the same halogeneous lamp obtained with ordinary straight gratings, while Fig. 13 (b) corresponds to the case of the circular gratings. The images at left were obtained in both cases by covering the 32 mm diameter lens with a slit 1 mm thick oriented orthogonally to the (X,Z) plane. Correspondingly, in images at right, the slit was located parallel to the X direction. A comparison of the Figs. 13 (a) and (b) shows that the circular gratings provide a significant reduction in astigmatism.

*4.2.2 Experimental verification of the image position*

In the present study, the numerical results of the model geometry were checked by measuring the image position with the focalization scale of a PENTAX-M reflex camera with an ASAHI 50 mm focal length objective. Different object positions were employed with the following input values: $\lambda_a$ = 500 nm, $\lambda_v$ = 700 nm, $Z_R$ = (126 ± 4) mm, and X = (37,0 ± 0,5) mm. The numerical results shown in Table II were obtained with Eqs. (2) and (4), using the Mathematica 5.0 software by Wolfram Research.

Table II: Comparison between the experimental and calculated values of the image distance.

| Z (mm) | $r_{exp}$ (mm) | $r_{cal.}$ (mm) |
|---|---|---|
| 301 ± 5 | 29 ± 3 | 25 ± 1 |
| 141 ± 5 | 23 ± 2 | 22 ± 1 |
| 90 ± 6 | 20 ± 2 | 20 ± 1 |
| 33 ± 9 | 10 ± 2 | 11 ± 2 |

Table II: Comparison between the experimental and calculated values of the image distance.

*4.2.3 Calculating the horizontal and vertical extension of the image*

4.2.3.1 Horizontal extension

In Figure 5, the two object points A and B with subscripts 1 or 2, indicating their different positions along the Z axis. Once again, the Eqs. (2) and (4) were used to determine the horizontal extension $\Delta Xi_{AB}$ of the image A', B', with the input values: $\nu$ = 658 lines/mm, $\lambda_a$ = 400 nm, $\lambda_v$ = 700 nm, and $Z_R$ = 126,5 mm. The calculated values of the horizontal image extension $\Delta Xi_{AB}$ are shown in Table III.

Table III: Calculated values of the horizontal image extension

| $Z_A = Z_B$ (mm) | $Xi_A$ (mm) | $Zi_A$ (mm) | $Xi_B$ (mm) | $Zi_B$ (mm) | $\Delta Xi_{AB}$ (mm) |
|---|---|---|---|---|---|
| 33 | -34.5 | 7.61 | -25.9 | 9.5 | 8.6 |
| 90 | -31.3 | 12.3 | -24.4 | 14.7 | 6.9 |
| 141 | -30.3 | 13.9 | -24.5 | 16.2 | 5.8 |

Table III: Calculated values of the horizontal image extension $\Delta Xi_{AB}$.

The chosen values for $X_A$ = 40 mm and $X_B$ = 30 mm corresponding to an object extension $\Delta X_{AB}$ = 10 mm, normal to the "optical axis" through the pinhole P.

4.2.3.2 Vertical extension

To calculate the vertical extension h' (Eq. 8) for two objects whose height is h located at positions $X_A$ =40 mm and $X_B$=30 mm, the chosen input values were: $\nu$ = 658 lines/mm, $Z_R$ is 126,5 mm and $\lambda_a$= 400 nm, and $\lambda_v$ = 700 nm .

The image extension h´ was determined with equations (6), (7), (8), using the software Mathematica 5.0. The derived results are shown in Table IV. The lengths on the drawing shown in Figure 5 preserve the scale of these results, i.e., they are drawn to exhibit the two cases of Table IV. In this figure, PI and PII identify the obligatory planes corresponding to the extremes of the object B.

Table IV: Vertical extension of the image

| $Z_A = Z_B$ (mm) | $Xi_A$ (mm) | $Zi_A$ (mm) | $Xi_B$ (mm) | $Zi_B$ (mm) | $h_A'$ (mm) | $h_B'$ (mm) |
|---|---|---|---|---|---|---|
| 33 | -35.6 | 7.4 | -27.2 | 9.2 | 8.6 | 8.6 |
| 90 | -32.2 | 12.0 | -25.5 | 14.3 | 7.8 | 8.1 |
| 141 | -31.0 | 13.6 | -25.4 | 15.8 | 7.5 | 8.0 |

Table IV: Vertical extension of the image when that of the object is h = 10 mm.

## 5. Conclusions

In the present work, it is demonstrated the feasibility of a new system capable of producing white light pseudoscopic and orthoscopic diffractive images. Calculations and measurements were shown to be in agreement. The choice of bidimensionally structured diffraction elements makes this system capable of projecting images which are pseudoscopic. Compared with previously reported systems [4,5], the present one allows the observation without astigmatism. This is advantageous since the images can be observed or photographed through a large aperture. In X-Ray optics distortionless images can be achieved no matter how large the bandwith.